\documentclass[preprint,aps,amsmath,byrevtex,prd,titlepage,
nofootinbib]{revtex4-1}

\usepackage{dcolumn}
\usepackage{amsmath}
\usepackage{amssymb}
\usepackage{bm}
\usepackage{hyperref}
\usepackage{graphicx}
\usepackage{slashed}

\newcommand{\beq}{\begin{equation}}
\newcommand{\eeq}{\end{equation}}
\newcommand{\eq}[1]{Eq.~(\ref{#1})}

\begin{document}

%\title {Hard Three-Loop Corrections to Parapositronium Energy Levels: Last Set of Diagrams with %Closed Electron Loop Insertions in Two-photon Annihilation Graphs}

\title{One More Hard Three-Loop Correction to Parapositronium Energy Levels}

\author {Michael I. Eides}
\altaffiliation[Also at ]{the Petersburg Nuclear Physics Institute,
Gatchina, St.Petersburg 188300, Russia}
\email[Email address: ]{eides@pa.uky.edu, eides@thd.pnpi.spb.ru}
\affiliation{Department of Physics and Astronomy,
University of Kentucky, Lexington, KY 40506, USA}
\author{Valery A. Shelyuto}
\email[Email address: ]{shelyuto@vniim.ru}
\affiliation{D. I.  Mendeleyev Institute for Metrology,
St.Petersburg 190005, Russia}
%\date{}

\begin{abstract}
A hard three-loop correction to parapositronium energy levels of order $m\alpha^7$ is calculated. This nonlogarithmic contribution is due to the insertions of one-loop photon propagator in the fermion lines in the diagrams with virtual two-photon annihilation. We obtained $\Delta E=0.03297(2)(m\alpha^7/\pi^3)$ for this energy shift.
\end{abstract}

%\pacs{36.10.Dr,12.20.Ds,31.30.jf,32.10.Fn}
%\keywords{hyperfine splitting}

%\preprint{UK/15-}

\maketitle

\section{Introduction}

Positronium, like hydrogen and muonium, is a loosely bound nonrelativistic two-particle  system. Two features make positronium special:  masses of the constituents are equal and the constituents can annihilate. The spectrum of positronium beyond the leading nonrelativistic approximation is significantly different from the hydrogen spectrum. This happens due to contributions of the annihilation diagrams and because the fine and hyperfine splittings have the same magnitude.  Theoretical research on positronium started in the second part of 1940s even before its experimental discovery \cite{pir47,ber49}, and was going ever after. As in other nonrelativistic systems there are two classes of corrections to energy levels, soft (and ultrasoft) and hard. Soft corrections originate from  a wide interval of virtual momenta below the electron mass, while only the virtual momenta of order of the electron mass are responsible for the the hard corrections. Soft corrections are usually logarithmically enhanced and in the case of positronium have the form of a double power series in $\alpha$ and $\ln\alpha$. Hard corrections can be calculated in the scattering approximation and in the case of positronium are simple series in powers of $\alpha$. Both the soft and hard corrections in positronium receive contributions from scattering and annihilation diagrams.

All corrections to hyperfine splitting (HFS) in positronium up to and including single-logarithmic terms of order  $m\alpha^7\ln\alpha$ were calculated before the end of the last or on the brink of the new millennium, see \cite{rjhill2001,kmay2001,bakaap2000} and brief reviews in \cite{bmp2014,af2014}. A new stage in the theory of positronium was initiated with calculation of the single-photon nonlogarithmic correction of order $m\alpha^7$ in \cite{bmp2014}. The ultrasoft contribution dominates this correction. Other soft and ultrasoft nonlogarithmic corrections of order $m\alpha^7$ remain unknown.

Many hard nonlogarithmic corrections of order $m\alpha^7$ were calculated recently in a rapid succession \cite{af2014,es2014r,apswf2014,apspl2015,es2015,akpprl2015,atw2016}. These corrections are generated both by the annihilation and non-annihilation diagrams.  Hard non-annihilation corrections are generated by seven gauge invariant sets of diagrams and are similar to the radiative and radiative-recoil corrections to HFS in muonium of orders $\alpha^2(Z\alpha)E_F$ and $\alpha^2(Z\alpha)(m/M)E_F$, respectively, see, e.g., \cite{egs2001,egs2007}. Corrections due to six gauge invariant sets of diagrams in muonium were calculated some time ago \cite{es2009prl,es2009pr,es2010jetp,es2013,es2014}. Contributions of the same six gauge invariant sets of diagrams  in positronium were obtained in \cite{af2014,es2014r,es2015}. These positronium  calculations in  \cite{es2014r,es2015}  were facilitated by our previous experience with the respective contributions in muonium.

Other hard nonlogarithmic corrections in positronium are generated by the diagrams with two, three, and four virtual annihilation photons \cite{atw2016}. There is one gauge invariant set of diagrams with four virtual annihilation photons, and three gauge invariant sets of diagrams with three virtual annihilation photons. Corrections due to the diagrams with four annihilation photons are currently unknown. Hard nonlogarithmic contributions of all diagrams with three annihilation photons were obtained in \cite{akpprl2015}.

\begin{figure}[h!]
\includegraphics
[height=2cm]
{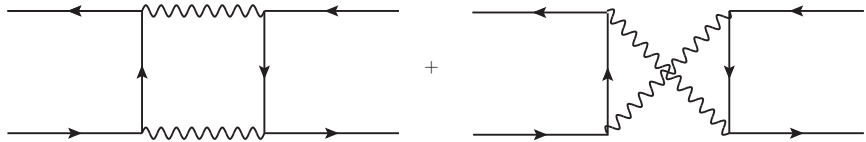}
\caption{\label{skelton}{Skeleton two-photon annihilation diagrams}}
\end{figure}

\begin{figure}[h!]
\includegraphics
[height=2cm]
{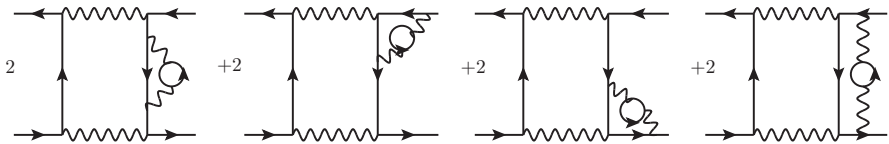}
\caption{\label{elfactpol}{Diagrams with polarization insertions in the radiative photon}}
\end{figure}

Hard corrections due to the diagrams with two annihilation photons are generated by seven gauge invariant sets of diagrams that are similar to the respective seven gauge invariant sets of non-annihilation diagrams in muonium and positronium \cite{es2014r,es2015}. All these diagrams can be obtained by two-loop radiative insertions in the skeleton diagrams with two annihilation photons in Fig.~\ref{skelton}. Contributions of five of these sets of diagrams were obtained in \cite{apswf2014,apspl2015,atw2016}.  Two sets of diagrams are still not calculated.  One of them is the set of diagrams with one-loop polarization insertions  in the radiative photon in Fig.~\ref{elfactpol} (the diagrams with the crossed annihilation photon lines are not shown explicitly in this figure). One more set of nineteen topologically different diagrams is obtained from the diagrams in Fig.~\ref{elfactpol} by deleting the polarization insertion from the radiative photon propagator and adding a second radiative photon insertion in the same fermion line.  Below we calculate hard nonlogarithmic correction of order $m\alpha^7$ generated by the gauge invariant set of diagrams in Fig.~\ref{elfactpol}.

\section{Contributions of Individual Diagrams}

\subsection{Skeleton Diagrams}

The skeleton diagrams with two-photon virtual annihilation in Fig.~\ref{skelton} generate hard corrections that contribute only to the shift of the parapositronium energy levels. These corrections were calculated long time ago \cite{kk1952}. We will briefly review the main features of the skeleton calculations following the recent discussion in \cite{atw2016}. These calculations will serve as a template for  calculations of the contributions of the diagrams in Fig.~\ref{elfactpol} below. The diagrams in Fig.~\ref{skelton} should be calculating in the scattering approximation and give contributions only to the parapositronium (spin zero) states with zero orbital momenta. The external electrons and positrons are on-shell and have zero spatial momenta. To obtain the contribution to the energy shift we project the amplitude on the spin zero states and multiply it by the  Coulomb-Schr\"odinger positronium wave function at the origin squared. The diagrams in Fig.~\ref{skelton} are both ultraviolet and infrared finite and give identical contributions to the energy level shift \cite{atw2016}. With account for all combinatorial factors the energy shift can be written as an integral over the loop four-momentum $k^\mu=(k_0,\bm k)$

\beq \label{leadingord}
\Delta E_s=\frac{m\alpha^5}{\pi}\int_0^\infty dk\int \frac{dk_0}{2\pi i}f_s(k_0,k),
\eeq

\noindent
where ($k=|\bm k|$)

\beq \label{fskel}
f_s(k_0,k)=-\frac{8m^2k^4}{[k_0^2-k^2+i0][(k_0-2m)^2-k^2+i0][(k_0-m)^2-k^2-m^2+i0]^2}.
\eeq

The principal feature of the annihilation diagrams in Fig.~\ref{skelton} is that they have imaginary part that arises because kinematics allows creation of two real photons. In agreement with the optical theorem this imaginary part contributes to the parapositronium decay width.  Existence of the imaginary part makes Wick rotation in the integral in \eq{leadingord} impossible, in the other case the diagram would be real. Considering  positions of the poles of the propagators in the box diagram we see that rotation in the plane of the complex $k_0$ without crossing one of the poles is impossible. The proper way to go is to calculate the integral over $k_0$ with the help of the residues, say in the upper half plane. The remaining one-dimensional integral over the magnitude $k=|\bm k|$ of the  three-dimensional loop momentum inherits a pole at $k=m+i0$ of one of the photon  propagators in the box diagram.  We use the Sokhotsky's formula to separate the real and imaginary parts of the momentum integral,  calculate both the real and imaginary momentum integrals analytically and reproduce the classic result \cite{kk1952}

\beq
\Delta E_s=\frac{m\alpha^5}{\pi}\left(\frac{1}{2}\ln2-\frac{1}{2}-\frac{i\pi}{4}\right).
\eeq

The pole in the one-dimensional integral survives in the diagrams with radiative insertions in Fig.~\ref{elfactpol} but its position in the general case is shifted. One still can calculate the real and imaginary parts of the respective integrals analytically in the same way as in the skeleton case.

Our strategy is first to calculate the contributions of the diagrams in Fig.~\ref{elfactpol} without the polarization operator insertions but with a finite radiative photon mass $\lambda$. We use the Feynman gauge in this calculations. In the limit of $\lambda\to0$ this calculation reproduces a well known contribution of order $m\alpha^6$ obtained in \cite{aab1993}, and serves as a test of our calculations. The integrals for the diagrams with polarization insertions in Fig.~\ref{elfactpol} are obtained from the respective diagrams without polarization insertion by the substitution $\lambda^2\to 4m^2/(1-v^2)$ followed by the integration over $v$ from zero to one with the weight $(\alpha/\pi)v^2(1-v^2/3)/(1-v^2)$, see, e.g., \cite{es2009}.

\subsection{Diagrams with Two-Loop Insertions}

%\paragraph
\subsubsection{Diagrams with Self-Energy Insertions}

We start with calculation of the diagrams with the self-energy insertions in Fig.~\ref{elfactpol}.  The well known renormalized self-energy operator has the form (see, e.g., \cite{blp}, we restored an exact dependence on the photon mass $\lambda$ below)

\beq  \label{mass-01}
\Sigma_R(p-k)=(m\gamma_0 -\slashed k-m)^2\frac{\alpha}{2\pi}\int_0^1 {dx} \int_0^x {dy}
\frac{mh_1(x,y)  - (m\gamma_0 -\slashed k+m)h_2(x,y)}{-k^2+2mk_0+a_1^2-i0},
\eeq

\noindent
where $m$ is the electron mass, $p=(m,\bm 0)$, $k=(k_0,\bm k)$,  and

\beq  \label{mass-h1h2}
h_1(x, y)= \frac{1+x}{y},\quad
h_2(x, y) =\frac{1-x}{y}\biggl[1-\frac{2(1+x)y}{x^2+\frac{\lambda^2}{m^2} (1-x)}\biggr],
\quad a_1^2(x, y) = \frac{m^2x^2+\lambda^2 (1-x)}{(1-x)y}.
\eeq

We  consider first the self-energy diagrams without polarization insertions but with a finite photon mass. We use the projector on spin zero (parapositronium) states (see, e.g., \cite{atw2016}) to get rid of the spinor structure and taking into account all combinatorial factors obtain an expression for the energy shift in the form similar to \eq{leadingord}

\beq
\Delta E_\Sigma(\lambda)=\frac{m\alpha^6}{\pi^2}\int_0^1 {dx} \int_0^x {dy}\int_0^\infty dk\int \frac{dk_0}{2\pi i}f_\Sigma(k_0,k),
\eeq

\noindent
where

\beq
\begin{split}
f_\Sigma(k_0,k)&=-8m^2k^4h_2(x,y)
[k_0^2-k^2+i0]^{-1}[(k_0-2m)^2-k^2+i0]^{-1}\\
&
\times[(k_0-m)^2-k^2-m^2+i0]^{-1}[k_0^2-k^2-2mk_0-a_1^2+i0]^{-1}.
\end{split}
\eeq

\noindent
Instead of the double fermion pole in the respective skeleton function $f_s(k_0,k)$ in \eq{fskel}, the function  $f_\Sigma(k_0,k)$ contains two simple poles. We again close the contour in the upper half-plane and use residues to calculate the integral over $k_0$. The real and imaginary parts of the integral over $k$ are separated with the help of the Sokhotsky's formula and calculated analytically. After integration over the Feynman parameters $x,y$ the integral at $\lambda\to0$ reproduces the infrared divergent contribution \cite{atw2016} of the one-loop self-energy insertion to the energy shift of order $m\alpha^6$.

The contribution to the energy shift of the self-energy diagrams with the vacuum polarization insertions in Fig.~\ref{elfactpol} requires one more integration

\beq
\Delta E_\Sigma=\frac{\alpha}{\pi}\int _0^1 dv\frac{v^2\left(1-\frac{v^2}{3}\right)}{1-v^2}\Delta E_\Sigma\left(\lambda\right)_{\bigl|\lambda=\sqrt{\frac{4m^2}{1-v^2}}}.
\eeq

\noindent
After numerical calculations we obtain

\beq \label{selftot}
\Delta E_\Sigma=(-0.028~960~328~(2)- 0.003~967~685~(2) i\pi)\frac{m\alpha^7}{\pi^3}.
\eeq

\subsubsection{Diagrams with Vertex Insertions}

To calculate the contribution of the diagrams with the vertex insertion in Fig.~\ref{elfactpol} we use the Feynman gauge expression for the one-loop vertex  with one virtual electron line and a finite photon mass, see, e.g., \cite{es2010} and references therein. This expression is too cumbersome  to cite it here. After some transformations we managed to represent the contribution of the vertex diagrams without polarization insertions but with a finite radiative photon mass and with  account of all combinatorial factors in the form

\beq \label{vetexinlamb}
\Delta E_V(\lambda)
=\frac{m\alpha^6}{\pi^2}\sum_{n=0}^2\int_0^1 dx\int_0^x dy\int_0^\infty dk\int\frac{dk_0}{2\pi i}
g_n(x,y)f_n(k_0,k),
\eeq

\noindent
where
$f_0(k_0,k)=f_s(k_0,k)$, see \eq{fskel},
\beq
\begin{split}
f_1(k_0,k)=&\frac{8m^2k^4}{[(k_0 - 2m)^2 - k^2+i0][(k_0 - m)^2 - k^2 - m^2+i0]^2[k_0^2 -
    k^2 - 2 mbk_0 - a^2+i0]},\\
f_2(k_0,k)=&\frac{8m^2k^4}{[k_0^2 - k^2+i0][(k_0 - m)^2 - k^2 - m^2+i0]^2[k_0^2 - k^2 -
    2 mbk_0 - a^2+i0]},
\end{split}
\eeq
and
\beq
\begin{split}
g_0(x,y)=&4\left(1-x-\frac{x^2}{2}\right) \frac{1}{m^2x^2+\lambda^2 (1-x)}
+\frac{2x^2}{\Delta_m},\\
g_1(x,y)=&\frac{2x^2}{\Delta_m}\left[y(1-y)-\frac{y(1-x)}{2}\right]
+2y(1-y)+2(x-y)(1-2y)+ \frac{2(1-x)^2}{2},\\
g_2(x,y) =&\frac{2x^2(1-x)y}{2\Delta_m} - \frac{2(1-x)^2}{2},\\
\Delta_m=&y(1-y)\left(2m^2b+a^2\right),\quad a^2=\frac{ m^2x^2+\lambda^2 (1-x)}{y(1-y)},\quad b = \frac{1-x}{1-y}.
\end{split}
\eeq

\noindent
We have adjusted the expression for the vertex in such way that only the function $f_0(k_0,k)$ contains both annihilation photon poles. As a result, only the terms in the integrand in \eq{vetexinlamb} that contain this function  generate both the real and imaginary contributions, the integrals of two other functions $f_1(k_0,k)$ and $f_2(k_0,k)$ are real. The momentum integrals in \eq{vetexinlamb} are calculated analytically like the momentum integrals in \eq{mass-01}, and the remaining integration over the Feynman parameters $x,y$ is done numerically.  At $\lambda\to0$ the integral for $\Delta E_V(\lambda)$ reproduces the infrared divergent contribution \cite{atw2016} of the one-loop vertex insertion to the energy shift of order $m\alpha^6$.

The contribution to the energy shift of the vertex  diagrams with vacuum polarization in Fig.~\ref{elfactpol} again requires one more integration

\beq
\Delta E_V=\frac{\alpha}{\pi}\int _0^1 dv\frac{v^2\left(1-\frac{v^2}{3}\right)}{1-v^2}\Delta E_V\left(\lambda\right)_{\bigl|\lambda=\sqrt{\frac{4m^2}{1-v^2}}}.
\eeq

\noindent
After numerical calculations we obtain

\beq \label{verttot}
\Delta E_V=(0.241~501~(2)-0.024~369~716~(2) i\pi)\frac{m\alpha^7}{\pi^3}.
\eeq

\subsubsection{Diagrams with Spanning Photon }

Calculation of the contribution of the diagrams in Fig.~\ref{elfactpol} with the spanning photon is the most cumbersome part of this work. It is well known that the respective diagrams without the one-loop polarization insertions in the photon propagator  contain a linear infrared divergence $m/\lambda$. This divergence is effectively cut off at the characteristic wave function momenta $\sim m\alpha$, which indicates that the respective diagrams contain a contribution of the previous order that should be subtracted. Insertion of the polarization operator in the spanning photon eliminates all infrared divergences. As a result the diagrams in Fig.~\ref{elfactpol} with the one-loop polarization insertions in the spanning photon are infrared finite and admit calculation in the scattering approximation.

Like in the case of the vertex we managed to represent the integral for the energy shift as a sum of convergent integrals and calculated the momenta integrals analytically. The remaining integrals over the Feynman parameters were done numerically and we obtained

\beq \label{spantot}
\Delta E_S=(-0.179~57~(2)-0.083~498~6\ldots -0.083~498~60~(3) i\pi)\frac{m\alpha^7}{\pi^3}.
\eeq

\noindent
Details of these calculations will be presented elsewhere.

\section{Summary of Results}

Collecting the results in \eq{selftot}, \eq{verttot}, \eq{spantot} we  obtain the total hard contribution to the parapositronium level shift of order $m\alpha^7$ generated by the diagrams in Fig.~\ref{elfactpol}

\beq \label{newresult}
\Delta E=(0.032~97~(2)-0.111~836~01~(3) i\pi)\frac{m\alpha^7}{\pi^3}.
\eeq

\noindent
The total hard contribution of order $m\alpha^7$ generated by the six gauge invariant sets of diagrams with two-photon annihilation is given by the sum of the correction in  \eq{newresult} and the results for the other five sets of annihilation diagrams calculated in \cite{apswf2014,apspl2015,atw2016}

\beq
\Delta E=0.901~67~(2)\frac{m\alpha^7}{\pi^3}=3.95940~(8)~\mbox{kHz}.
\eeq

\noindent
Analogous sum of hard contributions to HFS of the six gauge invariant sets of scattering diagrams   was calculated earlier \cite{af2014,es2014r,es2015}

\beq
\Delta E=-1.291~7~(1)\frac{m\alpha^7}{\pi^3}=-5.6720~(4)~\mbox{kHz}.
\eeq

\noindent
Combining these results with the hard three-photon annihilation contribution to the orthopositronium energy levels from \cite{akpprl2015}

\beq
\Delta E=2.621~6~(11)\frac{m\alpha^7}{\pi^3}=11.512~(5)~\mbox{kHz},
\eeq

\noindent
we obtain the hard contribution of order $m\alpha^7$ to HFS in positronium

\beq
\Delta E=0.428~(1)\frac{m\alpha^7}{\pi^3}=1.881~(5)~\mbox{kHz}.
\eeq

\noindent
This is still not a total hard contribution to HFS of order $m\alpha^7$. Five gauge invariant sets of diagrams remain unknown. These are two-photon exchange diagrams with insertions of two radiative photons in one and the same fermion line, a similar set of two-photon annihilation diagrams again with insertions of two radiative photons in one and the same fermion line, the set of diagrams with four-photon annihilation, and two sets of non-annihilation diagrams with recoil photons.

Calculation of these hard contributions as well as of soft corrections of order $m\alpha^7$ is the next goal of the positronium theory.

\acknowledgments

We are grateful to the referee for reminding us about two gauge invariant sets of uncalculated recoil diagrams that generate hard nonlogarithmic corrections which we forgot to mention in the original version of this paper.  This work was supported by the NSF grant PHY-1402593. The work of V. S. was also supported in part by the RFBR grant 16-02-00042.

\end{document}